%% file: OS_futility_interim.tex
\documentclass[12pt, a4paper]{article}

\input{commands.tex}

\begin{document}

\title{Integrating Phase 2 into Phase 3 based on an Intermediate Endpoint While Accounting for a Cure Proportion -- with an Application to the Design of a Clinical Trial in Acute Myeloid Leukemia}

\author{
    Kaspar Rufibach\thanks{Methods, Collaboration, and Outreach Group (MCO), Department of Biostatistics, Hoffmann-La Roche Ltd, Basel, Switzerland} \\
    Dominik Heinzmann\thanks{Oncology Biostatistics, Department of Biostatistics, Hoffmann-La Roche Ltd, Basel, Switzerland} \\
    Annabelle Monnet \footnotemark[2]\\
}

\date{\today}

\maketitle

\begin{abstract}
For a trial with primary endpoint overall survival for a molecule with curative potential, statistical methods that rely on the proportional hazards assumption may underestimate the power and the time to final analysis. We show how a cure proportion model can be used to get the necessary number of events and appropriate timing via simulation. If Phase 1 results for the new drug are exceptional and/or the medical need in the target population is high, a Phase 3 trial might be initiated after Phase 1. Building in a futility interim analysis into such a pivotal trial may mitigate the uncertainty of moving directly to Phase 3. However, if cure is possible, overall survival might not be mature enough at the interim to support a futility decision. We propose to base this decision on an intermediate endpoint that is sufficiently associated with survival. Planning for such an interim can be interpreted as making a randomized Phase 2 trial a part of the pivotal trial: if stopped at the interim, the trial data would be analyzed and a decision on a subsequent Phase 3 trial would be made. If the trial continues at the interim then the Phase 3 trial is already underway. To select a futility boundary, a mechanistic simulation model that connects the intermediate endpoint and survival is proposed. We illustrate how this approach was used to design a pivotal randomized trial in acute myeloid leukemia, discuss historical data that informed the simulation model, and operational challenges when implementing it.
\end{abstract}


\section{Introduction}
\label{intro}

Traditional clinical development of oncology drugs involves three distinct phases\cite{hunsberger_09}, each with its purpose and characteristic designs. Phase 1 serves to determine if a new drug is safe and the optimal dose for the next phase. Phase 2 aims to establish anti-tumor efficacy typically on a surrogate endpoint such as response proportion or progression-free survival (PFS). Finally, Phase 3 is typically a large randomized controlled trial (RCT) with an endpoint that is a direct measure of patient benefit, such as overall survival (OS), or an established surrogate thereof. However, in contemporary drug development, this paradigm is handled with increasing flexibility to trade-off some risks for a gain in speed\cite{hunsberger_09, chen_18}, in particular for disease settings with a high unmet medical need. As Chen et al.\cite{chen_18} note, ``Phase 2 proof-of-concept trials, which play a critical role in conventional drug development, are being skipped increasingly as a trade-off for speed''. Another approach to gain flexibility and speed is to dissolve the strict boundary between Phase 2 and Phase 3. Following the nomenclature in Hunsberger et al.\cite{hunsberger_09}, a trialist can {\it integrate Phase 2 into Phase 3}, meaning that a randomized Phase 3 is initiated directly after Phase 1 with a pre-planned futility interim analysis (IA) on an intermediate endpoint. This has been shown to substantially increase the probability of stopping early for futility when there is no treatment effect\cite{goldman_08}. Alternatively, Phase 2 can be {\it skipped}, meaning that the decision on futility at the IA is based on the Phase 3 primary endpoint. In both cases, if the Phase 3 trial is stopped at the IA, the final analysis of the trial would be performed. Thus, the sponsor finds itself basically in the same situation as if it had performed a randomized Phase 2 trial upfront. However, if the IA is passed, the Phase 3 is already well on its way.

When considering a time-to-event endpoint for a molecule potentially leading to cure, a proportion of patients will (at least within the duration of the clinical trial) never experience the event of interest. Major implications of such a scenario are that the survival functions do not exhibit proportional hazards (PH) over the entire time and simply assuming exponential survival functions for planning purposes is thus not optimal. This is because under non-PH the trial power depends on the exact shape of the survival function in both groups, event accumulation may taper off after a certain amount of follow-up, and the treatment effect could be reflected as a combination of improvement of the cure proportion and the delay of events for the uncured patients, see Sun et al.\cite{sun_18}. If a cure proportion is expected when planning a clinical trial, this has to be taken into account for computation of power (or alternatively, necessary number of events), choice of efficient statistical testing and analyses methods, and determination of trial duration.

The goal of this paper is to present a clinical trial design that integrates Phase 2 into Phase 3 in a disease setting where cure of a proportion of patients is possible. In Section~\ref{samplesizeos}, we discuss sample size simulation taking into account a cure proportion for a time-to-event endpoint, in our case OS. In Section~\ref{integrate}, conditions on the intermediate and primary endpoint that allow for integrating Phase 2 into Phase 3 are given. We describe a mechanistic simulation model that allows to connect the intermediate endpoint of tumor response to the primary endpoint OS, to be able to quantify the operating characteristics of any given futility interim decision boundary. The methodology is applied in Section~\ref{mirros} to illustrate the planning of the MIRROS trial, and operational implications are discussed. After discussing potential alternative designs to answer the relevant clinical question in Section~\ref{alternatives} we conclude the paper in Section~\ref{discussion}. Finally, some technical details that might be helpful when implementing the proposed design are deferred to the appendix.

\section{Sample size for a time-to-event endpoint with a cure proportion}
\label{samplesizeos}

As the proposed trial design neither allows for early stopping for efficacy nor for any adaptation based on the result of the IA (apart from potentially setting the Stage 2 sample size to 0 when stopping), the type I error of the final analysis on the primary endpoint is not inflated by adding the futility IA described in Section~\ref{integrate}. Section~\ref{opchar_IA} discusses how the overall power is affected by adding a futility IA. We thus first discuss powering a trial without considering the IA. Typically, in a two-arm RCT with a time-to-event endpoint, the null hypothesis
\bean
  H_0 \ : \ h_1 = h_2 \label{h0}
\eean of equal hazard (or survival) functions $h_i, i = 1, 2$ in the two groups of interest is assessed. For a pre-specified significance level $\alpha$, the number of events $d$ is determined such that a logrank test has power $1-\beta$ to reject the null hypothesis \eqref{h0} assuming a particular alternative hypothesis $H_1 : h_1 = \theta h_2$ is true, where $\theta$ is the hazard ratio (HR). Schoenfeld's formula\cite{schoenfeld_83} can then be used to compute $d$. If the PH assumption holds then this test is known to have maximum power\cite{cook_08}. Note that the logrank test is valid, in the sense of maintaining type I error, for any alternative, i.e. also when the PH assumption does not hold.

\subsection{The model}
\label{model}

The proposed design uses a mixture cure rate model\cite{boag_49}, see also Sun et al.\cite{sun_18} for a recent discussion in the clinical trial context. Let $S_i^*, i = 1, 2$ be the survival functions of the uncured patients for the control and experimental arm, respectively, and let $p_i, i = 1, 2$ be the proportions of patients cured. The survival functions in the proposed trial design can then be expressed as
\bean
  S_i(t) &=& p_i + (1-p_i)S^*_i(t), t \ge 0, \label{model_sample_size}
\eean with hazard functions
\bea
  h_i(t) &=&\frac{(1-p_i)f_i^*(t)}{p_i + (1-p_i)S_i^*(t)},
\eea with $f_i^*$ the density functions corresponding to $S_i^*$. Hence it is assumed that the experimental drug will act through both, increasing the proportion of cured patients and delaying events for those patients not cured. The ratio of hazard functions of treatment versus control group can be written as
\bea
  \theta(t) \ = \ h_2(t) / h_1(t) &=& \Bl(\frac{1-p_2}{1-p_1}\Br) \frac{f_2^*(t)}{f_1^*(t)} \Bl(\frac{p_1 + (1-p_1)S_1^*(t)}{p_2 + (1-p_2)S_2^*(t)}\Br).
\eea So, even in the simple model proposed by Sun et al.\cite{sun_18} where both $S^*_i$ are assumed to follow an exponential distribution with rates $\lambda_i$, the HR function $\theta(t), t\ge 0$ depends on time as soon as at least one of the $p_i > 0$, i.e. the PH ratio assumption does not hold in general in this simple cure proportion model. Absence of the PH assumption has several implications which we discuss in Section~\ref{trial_success}.

Although the literature generally refers to the above model as a ``cure rate model'' we prefer the term ``cure proportion'', as the parameters $p_i$ are indeed not rates, but proportions. Note that ``cure'' in the model for the $S_i$ must not necessarily refer to ``cure'' in the medical sense, but can also be ``long-term event-free''\cite{maller_92}. What constitutes ``long-term event-free'' depends on the disease indication.

\subsection{Planning a RCT through simulation}
\label{RCTplanning}

For simulating power or sample size in the above cure proportion model, one has to specify the quantities in Table~\ref{tab:ass1}. 

\renewcommand{\baselinestretch}{1.2}
\begin{table}[h]
\begin{center}
\begin{tabular}{l|c|c|c}
Quantity                              & Control arm & Treatment arm           \\ \hline
Survival function of uncured patients & $S^*_1$     & $S^*_2$                 \\ \hline
Cure proportion                       & $p_1$       & $p_2$                   \\ \hline \hline
$\#$patients recruited per month      & $n_{1j}$    & $n_{2j}$                 \\ \hline
Months of recruitment                 & \multicolumn{2}{c}{$j=1, \ldots, N$}   \\ \hline
Total $\#$patients recruited          & $n_1 = \sum_{j=1}^N n_{1j}$ & $n_2 = \sum_{j=1}^N n_{2j}$  \\ \hline
Drop-out rate per month               & $\tau_1$    & $\tau_2$                \\ \hline
\end{tabular}
\caption{Assumptions for simulation to compute necessary number of events $d$. The randomization ratio is defined as $r = n_2 / n_1.$}
\label{tab:ass1}
\end{center}
\end{table}
\renewcommand{\baselinestretch}{1.0}

It is assumed that for a given number of patients $n_{ij}$ recruited in arm $i$ in month $j$, their recruitment is uniformly distributed within that month. How to generate a random number from a survival function of a cure model is described in the appendix. For every simulated patient in arm $i$, we generate a drop-out time from an exponential distribution with arm-specific rate $\tau_i$. Assuming values for all the quantities in Table~\ref{tab:ass1}, the smallest number of events $d$ that gives power $1-\beta$ can then be determined by simulating $M$ clinical trials, administratively censor (on top of the censoring through drop-out) each of them at event numbers $d = d_0, \ldots, d_1$ ranging over a grid. Here, $d_0$ can e.g. be computed using Schoenfeld's formula assuming the overall target hazard ratio in a model without cure. Power is then estimated through the proportion of trials among the $M$ simulated that result in a logrank test rejecting $H_0$. The smallest value of $d$ that gives the targeted power is then declared the necessary number of events. In Sections~\ref{trial_success} and \ref{alternatives} potential alternative hypotheses with corresponding tests are outlined, and we discuss why these were not implemented.

One of the advantages of the PH assumption is that the power of the logrank test does indeed only depend on the specified significance level, number of events, and the assumed HR. That means for a chosen $d$ the targeted power is guaranteed if the assumed HR is correct. However, as discussed in Section~\ref{model}, the cure proportion model does in general not have the PH property. This implies that power does not only depend on the significance level and the chosen number of events $d$, but on the actual survival curves in both arms (not only the HR), recruitment, and dropout. We discuss operational implications of this feature in Section~\ref{opasp}.

To compute the necessary number of events for the cure proportion model above, a closed formula based on assumptions on effect sizes, recruitment, follow-up, and the distribution of censoring times has been proposed by Wang et al.\cite{wang_12}, making various assumptions on alternative hypotheses of interest. However, planning a trial through simulation offers several advantages: other quantities of interest besides power for a given $d$, e.g. timing of the final analysis or the probability of success, can easily be obtained based on the $M$ simulated trials. Also, uncertainty for these quantities can easily be asssessed based on the simulated trials. Furthermore, in MIRROS, although the PH assumption might not perfectly hold, interest was still in powering the trial for an overall logrank test, an alternative hypothesis that is not covered by Wang et al.

Recently, Sun et al.\cite{sun_18} discussed powering through simulation based on the cure proportion model above. They also advocated to use simulation to plan a RCT, and they described a simulation algorithm that is very similar to ours.

In the context of modelling the effect of a binary surrogate on a time-to-event endpoint, Abberbock et al.\cite{abberbock_18} also considered a similar mixture model as above, connecting the binary endpoint of pathological complete response to OS in early breast cancer. They did not assume a cure proportion and restricting attention to exponentially distributed time-to-event times.

\subsection{Trial success and quantification of effect}
\label{trial_success}

Since the cure proportion model will in general not fulfill the PH assumption, the standard unweighted logrank test will not be the most powerful test. The power loss compared to other tests depends on the assumed survival functions, i.e. the precise choice of the parameters in Table~\ref{tab:ass1}. In contrast to PH, power in a non-PH setup further depends on the censoring distribution, or more concretely, the actual accrual and dropout. Sun et al.\cite{sun_18} provide a small simulation study comparing different approaches. One finding is that even for cure proportions of 0.4 in the control and above in the experimental arm with hazard ratios of 0.7 or 0.8 in uncured patients, the power loss of a standard unweighted logrank against the considered optimized tests is generally modest. Since for regulatory submission purposes pre-specification of analyses is required, an a priori choice of a weight function (although proposals to adaptively choose the weights have been made as well) or a maximum follow-up time when considering restricted mean survival might be challenging. As a consequence, the logrank test seems to provide a good balance between universal applicability and power, even under deviations from PH in a cure proportion model. Note that this might not be true for other deviations from PH like crossing hazards or late separation.

As the (unweighted) logrank test is a valid test even under non-PH, a trial powered based on the above cure proportion model can be declared a success if the logrank test rejects $H_0$ in \eqref{h0}. A plot of the Kaplan-Meier estimates then provides the entire information that is contained in the trial sample data. However, trialists generally aim at summarizing the results in one or a small number of summary statistics. This facilitates understanding of the treatment effect by physicians and patients. Since the deviation from PH is anticipated to be small in MIRROS, the overall HR in all patients is still considered to be a meaningful effect quantifier. Furthermore, yearly estimates of the value of the survival functions in each group may be given.


Alternatively, estimates of the parameters in the cure proportion model in Section~\ref{model} could be used to quantify the effect, see e.g. Peng and Dear\cite{peng_00} or Tsodikov et al.\cite{tsodikov_03}.

\section{Integrate Phase 2 into Phase 3 and operating characteristics}
\label{integrate}

Depending on the interplay between recruitment and how quickly events accrue for the primary endpoint, the uncertainty around an effect estimate based on the primary endpoint at the futility IA may be substantial. Hence, skipping Phase 2 entirely as described in Section~\ref{intro} may not be optimal. Instead, the futility IA decision can be based on one or multiple suitable intermediate endpoint(s). Since the intermediate endpoint is only used for a futility decision and not to potentially establish efficacy at an IA, full surrogacy of the intermediate endpoint is not required. Rather, it needs to be sufficiently associated with the primary endpoint\cite{goldman_08}. The meaning of ``sufficiently associated'' depends on various aspects, including disease setting and how endpoints are linked with ultimately, the operating characteristics associated with the interim decision based on the selected intermediate endpoint.

For a model with OS as primary and PFS as intermediate endpoint, Hunsberger et al.\cite{hunsberger_09} discuss that under the assumption of no effect (``global null'', no effect on PFS and OS), integrating Phase 2 into Phase 3 can substantially reduce development time and number of patients compared to skipping Phase 2, and is obviously comparable to the traditional paradigm of performing a randomized Phase 2 trial followed by a randomized Phase 3. Reason being that futility monitoring on PFS entails more events earlier compared to OS.

If assuming that the experimental treatment has an effect on both, PFS and OS (``global alternative''), findings are that integration of Phase 2 leads to development time and number of patients comparable to skipping Phase 2, but which are in turn substantially lower compared to the sequencing approach. So, integrating Phase 2 into Phase 3 does not lose in either scenario but outperforms each potential alternative in turn. While the literature focuses on an intermediate endpoint of PFS for a primary endpoint of OS, the conclusions remain valid also in the scenario we describe below with a response-type intermediate endpoint.

Once an interim futility boundary has been chosen, quantification of the associated risk can be done via estimating probabilities of a correct or wrong decision, assuming either the experimental drug does not work or works according to an assumed effect. These probabilities are generally called the {\it operating characteristics} of the futility IA.

If the interim decision is based on the primary endpoint, trialists typically look at conditional power, i.e. updating the initial power calculation with the interim result. Such an update is even possible if the sponsor remains blinded to the detailed IA results\cite{rufibach_14}.

To quantify the above probabilities for the interim decision based on an intermediate endpoint, one typically has to resort to a simulation approach to model the association between the intermediate and the primary endpoint. This can be achieved  by using a mechanistic model explicitly linking the endpoints, as described in Section~\ref{integrate_mod} or Abberbock et al.\cite{abberbock_18}.

\subsection{Choice of intermediate endpoint}
\label{inter}

Given being a complete responder (complete response, CR) is a necessary condition for being a long-term survivor (referred to as ``cure''), response may be considered ``sufficiently associated'' to OS and thus seems a reasonable choice for the intermediate endpoint in a setting with OS the primary endpoint. In what follows, we will thus focus on this scenario of OS being primary and response (or another binary endpoint) being the intermediate endpoint, and we will illustrate how operating characteristics can be estimated in this scenario.


\subsection{Mechanistic simulation model to connect response and OS}
\label{integrate_mod}

\begin{figure}[htb]
\begin{center}
\setkeys{Gin}{width=1\textwidth}
\includegraphics{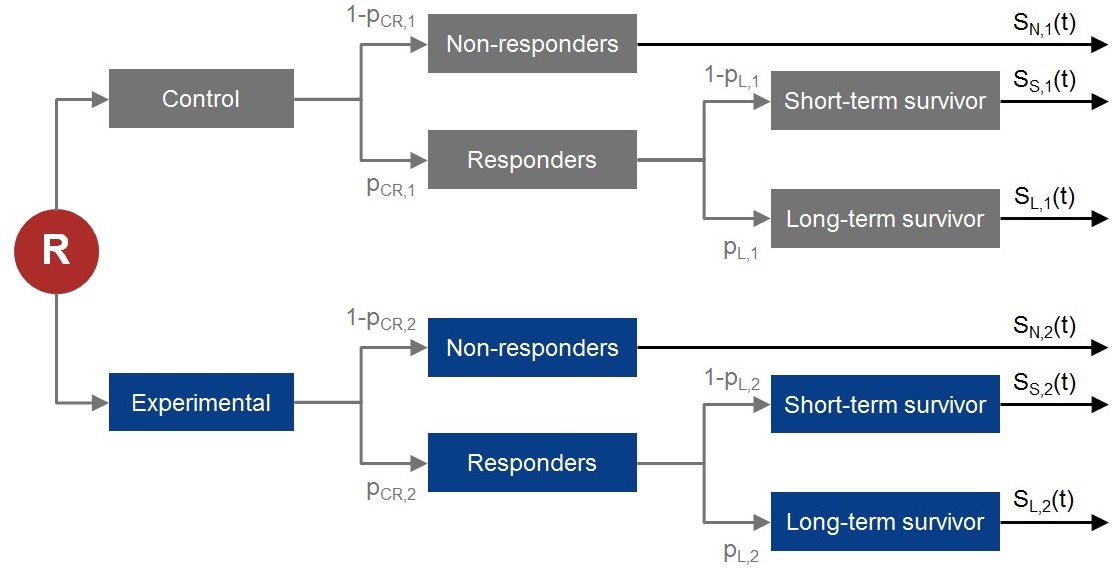}
\caption{Flowchart of mechanistic simulation model. ``R'' stands for ``Randomization''.}
\label{fig:mechanistic}
\end{center}
\end{figure}

For the proposed design, CR is only a necessary but not a sufficient condition for cure. Hence the mechanistic model in Figure~\ref{fig:mechanistic} makes a distinction between patients who respond short- or long-term, implying three groups of patients in each of two arms of the RCT: non-responders, short-, and long-term responders. To be able to simulate from this model, one has to determine the probabilities for a patient to be in one of these six groups, and the survival function in each group. We summarize these quantities in Table~\ref{tab:ass2}, complemented by assumptions on recruitment and drop-out identical to those in Table~\ref{tab:ass1}.

\renewcommand{\baselinestretch}{1.2}
\begin{table}[h]
\begin{center}
\begin{tabular}{l|c|c|c}
Quantity                              & Control arm           & Treatment arm            \\ \hline
Survival function of non-responders   & $S_{\text{N}, 1}$     & $S_{\text{N}, 2}$                  \\ \hline
Probability to have CR                & $p_{\text{CR}, 1}$    & $p_{\text{CR}, 2}$                  \\ \hline  
Probability to be long-term responder | CR & $p_{\text{L}, 1}$   & $p_{\text{L}, 2}$                  \\ \hline
Survival function of short-term responders & $S_{\text{S}, 1}$     & $S_{\text{S}, 2}$                  \\ \hline
Survival function of long-term responders & $S_{\text{L}, 1}$     & $S_{\text{L}, 2}$                  \\ \hline \hline
$\#$patients recruited per month      & $n_{1j}$    & $n_{2j}$                 \\ \hline
Months of recruitment                 & \multicolumn{2}{c}{$j=1, \ldots, N$}   \\ \hline
Total $\#$patients recruited          & $n_1 = \sum_{j=1}^N n_{1j}$ & $n_2 = \sum_{j=1}^N n_{2j}$   \\ \hline 
Drop-out rate per month               & $\tau_1$    & $\tau_2$                \\ \hline
\end{tabular}
\caption{Assumptions for simulation related to the mechanistic simulation model.}
\label{tab:ass2}
\end{center}
\end{table}
\renewcommand{\baselinestretch}{1.0}

This model is an extension of the simple cure proportion model introduced in Section~\ref{samplesizeos} and the corresponding survival functions can be written as:
\bean
  \bar{S}_i(t) &=& p_{CR,i}\Bl(p_{L,i} + (1-p_{L,i})S_{S,i}(t)\Br) + (1-p_{CR,i})S_{N,i}(t), t \ge 0, \label{model_cr_os}
\eean with $i=1,2$ for the control and experimental arm, respectively.

\subsection{Simulating from the mechanistic model}
\label{integrate_sim}

To simulate from the mechanistic model as introduced in Section \ref{integrate_mod}, one can extend the cure proportion simulation scheme outlined in Section \ref{samplesizeos}. To this end, one has to simulate survival times for virtual patients in the six groups detailed in Figure~\ref{fig:mechanistic} according to the assumptions from Table~\ref{tab:ass2}, where group sizes are determined through the probabilities to be a non-, short-, or long-term responder.

Making assumptions about the quantities in Table~\ref{tab:ass2} reflecting either a scenario where the experimental drug works just as the control, or improves on the control, operating characteristics as described in Section~\ref{inter} can be computed via simulation. Accordingly, the futility interim boundary can be selected such that there is an appropriate trade-off between the error probabilities to stop the trial early if the drug works (false-negative) and continue after the interim if the drug does not work (false-positive).

\subsection{Separate models to power the trial and evaluate operating characteristics}
\label{os_vs_mechanistic}

It remains to be clarified why we chose to have two separate models for powering a trial and adding a futility IA. In general, sample size determination is a trade-off between using a simple (maybe even simplifying) and robust model making few assumptions and a more sophisticated model that typically necessitates to assume more input quantities, compare Tables~\ref{tab:ass1} and \ref{tab:ass2}. The assumptions that inform any such model all come with uncertainty, so that the ``gain'' in precision by using a more sophisticated model might be offset by uncertainty introduced by the additional assumptions.

For protocol writing and getting informative health authority feedback in the early stage of trial development, it is helpful to have a reasonably simple and robust method to determine the sample size. The cure proportion model in Section~\ref{samplesizeos} meets these criteria and can be used to derive the necessary number $d$ of events to power the trial. To this end, only the assumptions in Table~\ref{tab:ass1} are required.

If one accepts the power loss induced by adding a futility IA, then powering and adding a futility IA can be done independently of trial powering, i.e. after $d$ had been determined. Often, certain trial assumptions such as futility boundaries remain under discussion even after the trial started, e.g. due to evolving external evidence or deviations from the initial recruitment assumptions. Having two separate models allows for fine-tuning of the IA (e.g. the IA boundary) without having to update the sample size part. Of course, any approach that models the IA needs to be able to reproduce the sample size computation. To achieve this, for a given $i = 1, 2$, input parameters for the survival functions $S_i$ in \eqref{model_sample_size} and $\bar{S}_i$ in \eqref{model_cr_os} need to be chosen such that the resulting survival functions match. To this end, first choose $p_\text{CR, i}, p_\text{L, i},$ and $p$ such that $p = p_\text{CR, i} \cdot p_\text{L, i}$. Then, calibrate $S_i^*, S_{S, i},$ and $S_{L, i}$ such that $S_i(t) = \bar{S}_i(t)$ for $t \ge 0$, where calibration either means choose these quantities based on historical data and/or make assumptions for them.

\section{Case study: MIRROS}
\label{mirros}



The design that we propose was used to design a real clinical trial in acute myeloid leukemia (AML).

MIRROS\cite{mirros} is a Phase 3 multicenter, double-blind, randomized, placebo-controlled trial of Idasanutlin in combination with Cytarabine compared with Cytarabine and placebo in relapsed-refractory (R/R) AML. The trial was initiated based on promising Phase 1 data and high unmet medical need for patients with R/R AML\cite{reis_16}. For details on the mechanism of action of Idasanutlin we refer to Tovar et al.\cite{tovar_06} and Vassilev et al.\cite{vassilev_07}.

At the time of the design of MIRROS, OS was the preferred endpoint. Although some patients with AML benefit from chemotherapy-based regimens, these therapies are rarely curative with the exception of stem cell transplant\cite{forman_05}. For this reason, it appears sensible to incorporate a cure proportion when powering the trial.

Several aspects contributed to the decision to integrate Phase 2 into Phase 3 in MIRROS. First and foremost, the very high unmet medical need in R/R AML, supporting accelerated development. Second, early phase results with Idasanutlin were considered encouraging by the sponsor. 
Finally, there exists a suitable intermediate endpoint (CR, binary) related to OS for the integration, see Section~\ref{inter}. Here, we omit the details of the clinical assessment of CR and just mention that
response-type endpoints have been previously considered suitable as intermediate endpoints\cite{hunsberger_09}. Details of the IA are provided in Section~\ref{opchar_IA}. Here, we only mention that the timing of the IA was determined through a pre-specified number of patients that are evaluable for CR at the IA, namely 120. This number was considered suitable for making the interim decision and corresponds to the sample size of a typical randomized Phase 2 trial.

A point to clarify is why we have not based the interim decision on OS, i.e. skipping Phase 2 entirely. Advantages to use OS for futility would have been that (1) no discussion would have been needed on how well the intermediate is associated with the primary endpoint, (2) operating characteristics of the interim could easily be evaluated using conditional power considerations. However, overriding disadvantages using OS for a futility assessment would have been:
\bi
\item If Idasanutlin works as assumed under the alternative used to power the trial (denoted by $H_1$), then we expect parts of the OS effect to be driven by cured patients. However, at the time of the IA, survival function estimates would not yet have been constant, i.e. only a small number out of the 120 IA patients would have been under observation for long enough for them to be considered cured.
\item Early deaths are known to occur on the control arm, and it is anticipated that the proportion of early deaths could be slightly higher in the experimental arm due to potential synergistic toxicities between Cytarabine and Idasanutlin. This bears the risk that an interim decision based on OS might be confounded by higher number of early deaths in the experimental arm.
\ei

To conclude this subsection, we mention that in order to maintain trial integrity and avoid bias, the sponsor would remain blinded to the IA result. Only an independent data monitoring committee (iDMC) would see unblinded results and provide a recommendation to the sponsor, based on criteria that are summarized in an iDMC charter.

\subsection{Assumptions for sample size model based on historical data}
\label{ass_ss}

In order to simulate from the cure proportion model described in Section~\ref{samplesizeos} we need to make assumptions about all the parameters in Table~\ref{tab:ass1}:
\bi
\item For the hypothesis test \eqref{h0}, a two-sided significance level of $\alpha = 0.05$ was assumed.
\item The randomization ratio $r$ was set to 2, implying that twice as many patients were randomized to the experimental compared to the control arm. Reasons for setting $r=2$ were optimization of the safety database for Idasanutlin as well as experience with the novel treatment\cite{dumville_06}, and increase in odds for patients to be on the experimental arm.
\item The survival functions of uncured patients, $S_i^*$, were both assumed to follow an exponential distribution, with rates $\lambda_1^* = 0.131$ and $\lambda_2^* = 0.101$. These rates correspond to medians of $m_1^* = 5.304$ and $m_2^* = 6.880$ months, respectively.
\item The cure proportions are assumed to be $p_1 = 0.080$ and $p_2 = 0.161$, respectively. The rationale for assuming these values is given in Section~\ref{ass_ia}.
\ei
With these choices, the survival function $S_1$ in the control arm of MIRROS has a six month median survival. This median appeared sensible, for the following reasons:
\bi
\item In two comparable Phase 3 trials, CLASSIC I\cite{faderl_12} and VALOR \cite{ravandi_15}, OS medians were slightly above six months in their Cytarabine plus placebo arms.
\item However, enrolment of poor prognosis patients with second R/R disease, in addition to first relapse, was allowed in MIRROS, which led to assuming a slightly lower median OS.
\ei
For the experimental arm, median OS is assumed to be 9 months. A three-month median OS improvement from 6 to 9 months was considered to be clinically meaningful for these patients for which no standard efficacious treatment options exist. Under exponentiality, i.e. without the cure proportion, these medians would correspond to a HR of 6.0 / 9.0 = 0.67. The critical value of the two-sided logrank test on the HR scale corresponding to these assumptions would be 0.78. This quantity specifies at which effect size the $p$-value would be equal to 0.05, i.e. the logrank test would ``just be significant''. This is typically called the {\it minimal detectable effect size}.

\begin{figure}[h!tb]
\begin{center}
\setkeys{Gin}{width=0.85\textwidth}
\includegraphics{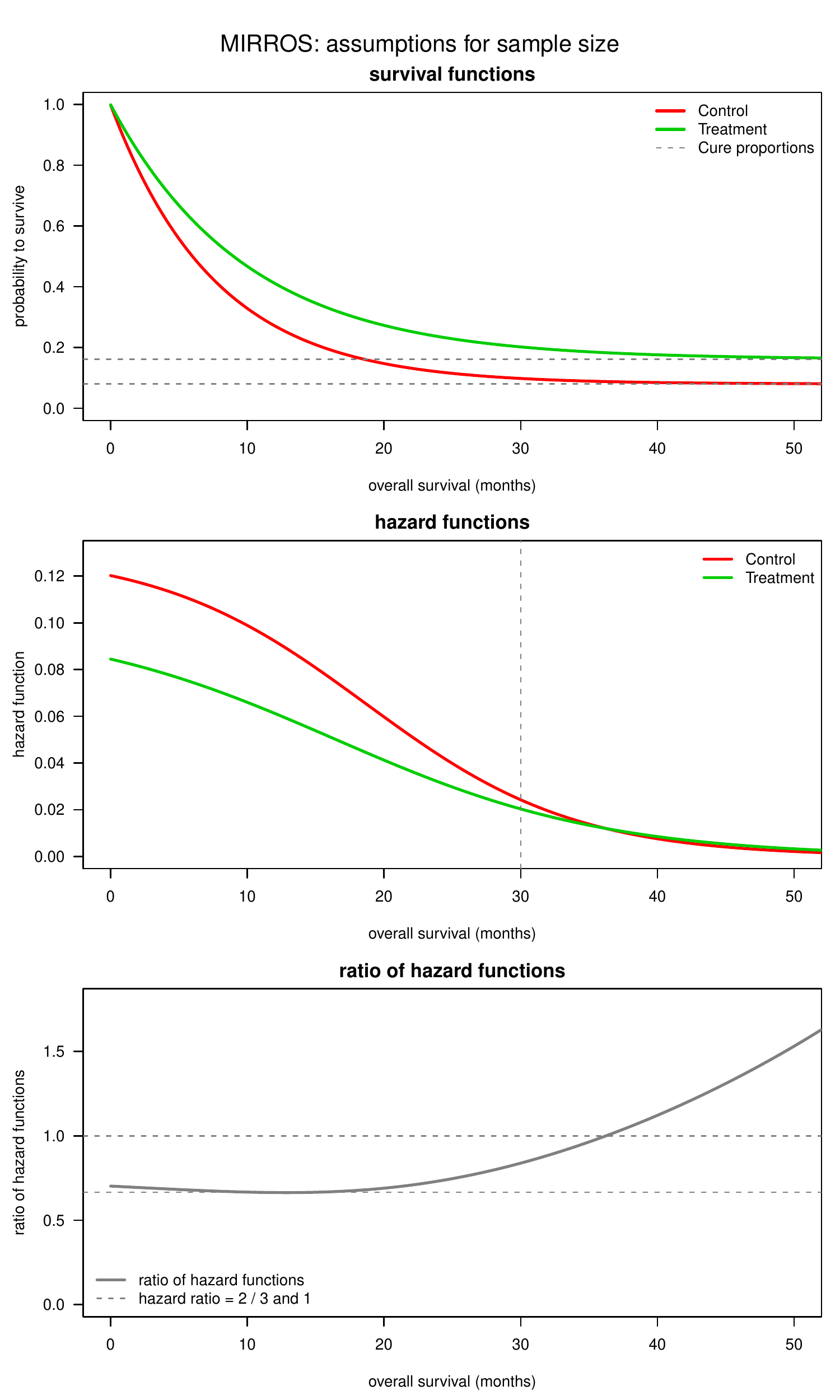}
\caption{Assumptions on OS survival functions $S_i^*$. Horizontal grey lines are assumed cure proportions.}
\label{fig:assumptions_os}
\end{center}
\end{figure}

With these choices for $S_i^*$ and $p_i$, the resulting survival $S_i$ and hazard $h_i$ functions in the control and experimental arm are displayed in the top row of Figure~\ref{fig:assumptions_os}. We note that the hazard functions are getting very close to 0 at approximately three years after randomization. This timepoint can thus be interpreted as a lower bound for when a patient can be considered ``cured''. This is in line with what is reported in de Lima et al.\cite{lima_97}, who consider AML patients to be potentially cured if they remain in first or second CR for at least three years, with a failure rate below 0.1 thereafter (Table 1 in de Lima et al\cite{lima_97}).

As discussed in Section~\ref{intro}, in the presence of non-PH, the power (or the necessary number of events) also depends on the accrual. Statistically, we can simply make assumptions on $n_{j} = n_{1j} + n_{2j}, j = 1, \ldots, N$. In MIRROS however, the mechanism of action of Idasanutlin requires activation of p53. This is expected in p53 wild-type AML and possible in p53 mutations that retain functionality of the protein. For that reason, the primary efficacy endpoint for which the trial is powered is OS in p53 wild-type patients, and it is assumed that these make up $85\%$ of patients recruited in the trial\cite{forbes_08, rucker_12}. Since determining the patient's p53 status through a central laboratory would have delayed randomization by several weeks, and given the rapid progression of AML, p53 status was not used for stratification. OS in the overall trial sample is a secondary endpoint in a hierarchical testing strategy for which no power computation is performed.

Overall recruitment is assumed to be $\tilde{n}_{\cdot j} := \tilde{n}_{1j} + \tilde{n}_{2j} = 12$ patients / month for 15 months, followed by 17 patients / month for another 15 months until a total of 440 patients have been recruited. At trial planning, it was anticipated that the interim decision would be made after about 15 months and if the trial continued, recruitment would be increased through opening of additional trial sites.

Now, in every month $j$ we assume that $n_{\cdot j}$ is randomly drawn from a Binomial distribution with $\tilde{n}_{\cdot j}$ experiments and probability 0.85. In turn, the number of patients in the control arm $n_{1j}$ are then again drawn from a Binomial with $n_{\cdot j}$ experiments and probability $1 / 3$, and finally $n_{2j} = n_{\cdot j} - n_{1j}$. Once the number of patients recruited per month has been determined, arrival times within a month are generated using a Poisson process.

The annual dropout rate was assumed to be 0.05 in both arms, i.e. $\delta = \delta_1 = \delta_2 = 0.05$. From this we get $\tau = \tau_1 = \tau_2 = 0.0043$, see the appendix. This assumption is based on the Sponsor's internal dropout data typically observed in oncology clinical trials.

\subsection{Sample size}
\label{SampleSize}

For sample size computation, a power of $85\%$ was assumed to detect the effect induced through the assumptions in Section~\ref{ass_ss}. With this, $M = 100000$ trials were simulated, for a range of assumed number of events $d = d_0, \ldots, d_1$ as discussed in Section~\ref{RCTplanning}. As can be seen from Table~\ref{tab:MIRROSsize}, 275 events provide the targeted power of $85\%$. According to Schoenfeld's formula, a design without assuming a cure proportion for a hazard ratio of 6.0 / 9.0 = 0.67 would need 246 events. So, accounting for the cure proportion in MIRROS increases the required number of events by $12$\%, compared to a standard design with the same median OS assumptions (Row 2 in Table~\ref{tab:MIRROSsize}). Or, using 246 events with the MIRROS assumptions would give a $4\%$ lower power ($81.0$\%). For the template design using an exponential assumption without a cure proportion, it would take $29.2$ months (median over 100000 simulations) until the targeted number of events would be reached. Accounting for the cure proportion in MIRROS increases this trial duration by $9.6$ to $38.8$ months.

\renewcommand{\baselinestretch}{1.2}
\begin{table}[h]
\begin{center}
\begin{tabular}{c|l|c|c|r|r|r|r|r}
Scenario & Assumption & $S_1^{-1}(0.5)$ & $S_2^{-1}(0.5)$ & $p_1$ & $p_2$ & $d$ & power & time \\ \hline
1 & MIRROS (non-PH)     & 6.0 & 9.0 & 0.080 & 0.161 & 275 & 0.852 & 38.8\\ \hline

2 & PH, no cure     & 6.0 & 9.0 & 0 & 0 & 246 & 0.858 & 29.2\\ \hline

3 & MIRROS (non-PH) with& 6.0 & 9.0 & 0.080 & 0.161 & 246 & 0.810 & 33.7\\

& $\#$events for PH, no cure     & & & & & & & \\ \hline
\end{tabular}
\caption{Necessary number of events and power for three sets of assumptions. The last column {\it time} shows the median time until the necessary number of events is reached, in months, with median taken over the $100000$ simulation runs. For Scenarios 1 and 3 sample size $d$ was computed using simulations, for Scenario 2 via Schoenfeld's formula.}
\label{tab:MIRROSsize}
\end{center}
\end{table}
\renewcommand{\baselinestretch}{1.0}

The bottom two rows of Figure~\ref{fig:assumptions_os} show the hazard functions that correspond to the assumptions of Scenario 1 in Table~\ref{tab:MIRROSsize}, together with their ratio. Up to about 25 months the PH assumption is approximately met. Only afterwards, the survival functions start to approach their respective cure proportion and we see a relevant deviation of the PH assumption. This is in line with the above finding that in fact, the power loss of the logrank test induced through the violation of this assumption remains small.

\subsection{Assumptions for the interim analysis model}
\label{ass_ia}

For the more general IA model we need to make assumptions about all the parameters in Table~\ref{tab:ass2}.

The key feature of the intermediate endpoint in MIRROS, CR, is that it is binary. A CR proportion $p_\text{CR, 1} = 0.16$ was assumed in the control arm, since CR rates of 17.8\% and 16.0\% were observed at the end of induction in the corresponding arm of CLASSIC I and VALOR respectively, so a lower CR proportion of 0.16 was chosen to take into account the inclusion of patients with second R/R disease in MIRROS. The overall proportion of ``cured'' patients, or at least long-term survivors among patients with CR, was assumed to be 0.5 for both, $p_\text{L, 1}$ and $p_\text{L, 2}$. This is higher than in the prospective randomized VALOR and CLASSIC I trials, where 30\% and 20\% of CR patients overall underwent transplantation, because MIRROS specifically targets patients who are suitable to undergo transplantation if achieving CR. The decision to transplant or not a patient is multi-factorial and depends on a myriad of criteria, the most important ones being minimum residual disease, patient fitness, donor compatibility, or remaining toxicities from previous treatment\cite{rashidi_18}.

To meaningfully increase the cure proportion, a large increment in CR was targeted. An odds ratio of OR = 2.5 was assumed, resulting in a CR proportion in the experimental arm of
\bea
  p_{\text{CR}, 2} &=& \text{OR} \cdot o_{\text{CR}, 1} / (1 + \text{OR} \cdot o_{\text{CR}, 1}) \ = \ 0.323,
\eea where $o_{\text{CR}, 1} = p_{\text{CR}, 1} / (1 - p_{\text{CR}, 1})$ are the odds for CR in the control arm.

Taking the assumptions on CR and long-term responding proportions together, assumed cure proportions were thus $p_1 = 0.160 \cdot 0.5 = 0.080$ and $p_2 = 0.323 \cdot 0.5 = 0.161$ in the control and experimental arm, respectively, matching the assumptions of the reduced model in Section~\ref{ass_ss}.

Finally, $S_{N,i}, S_{S, i},$ and $S_{L, i}$ were all again assumed to follow an exponential distribution. In the control arm, medians were chosen as 5.13 and 7.5 months, respectively. These assumptions were inspired by reading-off medians from Figure 3B in Kurosawa et al.\cite{kurosawa_10}. In this figure, displayed estimated survival functions can approximately be identified, from bottom to top, with $S_{N, 1}, S_{S, 1},$ and $S_{L, 1}$. The median for the green curve, which corresponds to $S_{S, 1}$, as read of from the plot is about 7.5 months. For the grey curve corresponding to $S_{N, 1}$ we shortened the median from the figure slightly.

To get the experimental arm medians, an improvement of a hazard ratio of 0.8 on top of the control arm medians was assumed. Finally, for the long-term survivors in both arms a deterministic large survival time was imputed in the simulations. With these choices, the resulting survival functions were calibrated as described at the end of Section~\ref{os_vs_mechanistic}, i.e. $S_i = \bar{S}_i, i = 1, 2.$

\subsection{Operating characteristics of the futility interim analysis}
\label{opchar_IA}

\begin{figure}[h!tb]
\begin{center}
\setkeys{Gin}{width=0.85\textwidth}
\includegraphics{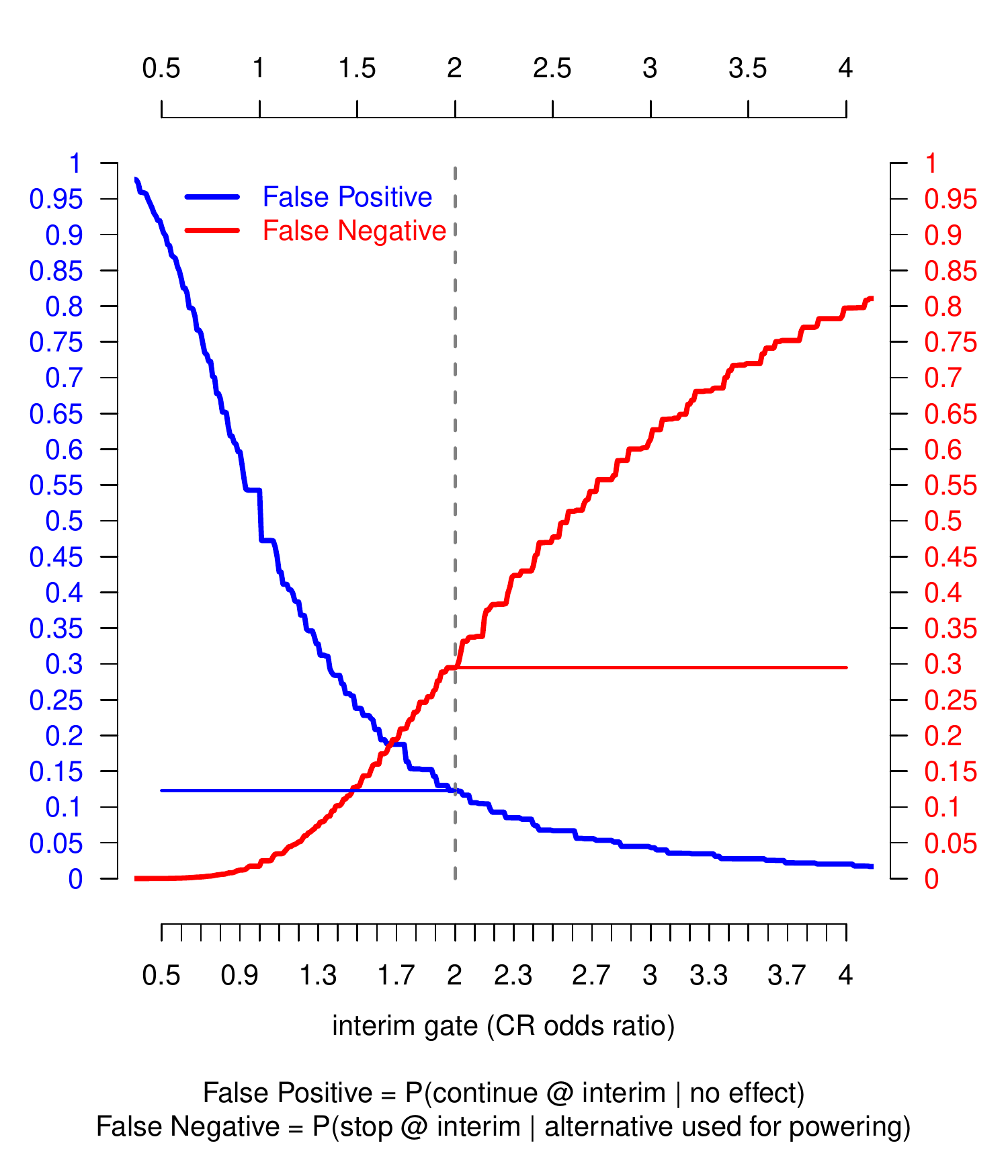}
\caption{False-positive and false-negative probabilities for a range of futility boundaries.}
\label{fig:fp_fn}
\end{center}
\end{figure}

To summarize, integration of Phase 2 into Phase 3 in MIRROS was based on a futility IA using the intermediate endpoint CR. Timing thereof was when CR data of 120 patients had become evaluable, and hence approximately mimicking the size of a randomized Phase 2 trial. The odds ratio has been selected as effect quantifier for the intermediate endpoint CR, such that given the selected interim boundary $\intb$, an observed odds ratio above that boundary would indicate that the trial should pass the IA. In line with the strategy outlined in Section~\ref{opchar_IA}, $\intb$ can be selected by targeting an appropriate trade-off between false-positive and false-negative error probabilities. Those error probabilities are presented in Figure~\ref{fig:fp_fn} for different choices of $\intb$. A value of $\intb = 2$ was considered a sensible trade-off between the error probabilities of false-negative 0.123 and false-positive 0.295.

With $\intb=2$, the relatively high false-negative error probability (as compared e.g. to recommendations by Gallo et al.\cite{gallo_14}) indicates that a high interim bar was set, meaning that not stopping the trial at the IA increases confidence in the experimental arm. Adding a futility IA to a trial comes at the cost of reduced power. In MIRROS, the power for the logrank test at the final analysis drops from 0.85 assumed for sample size computation in Section~\ref{SampleSize} to 0.63 when taking into account the futility IA. This power loss can easily be quantified from our mechanistic simulation model, further illustrating the usefulness of such a setup.

The probability to stop if the experimental performs as the control arm amounts to 1 - 0.123 = 0.877. Given that this Phase 3 trial was initiated based on Phase 1 data only, this is precisely the risk mitigation that was targeted, i.e. make it hard for the trial to pass the interim in case the drug does not work. The two error probabilities also have to be seen in context of recommendations for type I and II error probabilities in randomized Phase 2 trials. For example, Rubinstein et al.\cite{rubinstein_09} explore features of randomized Phase 2 studies with probabilities of type I and/or II errors of up to 0.2.

Note that $\intb=2$ would be used by the iDMC in their decision framework (see Section~\ref{opasp}). However, as Jitlal et al.\cite{jitlal_12} emphasize, stopping a trial early is a crucial decision to be made between the iDMC and the sponsor. Hence additional futility criteria around safety and early deaths have been implemented for the iDMC to enable a robust assessment on the benefit-risk ratio of the experimental treatment.

\subsection{Operational aspects}
\label{opasp}

Basic features of flexible designs including IAs, and related operational aspects, have been discussed elsewhere \cite{fda_interim, fda_adaptive, ema_adaptive}. Here, the focus is on a few operational considerations pertaining to the futility IA in MIRROS.

To maintain blinding and ensure trial integrity, the decision whether to stop the trial at the futility IA was based on the recommendation of an iDMC. The iDMC charter summarized the criteria for futility and safety.

Based on the assumptions, an interim decision was expected approximately 15 months after the first patient was randomized. This number is the sum of the timepoint when the 120th wild-type patient is randomized, duration for assessing response, and iDMC process.

At the time of the iDMC recommendation, it was expected that of the 440 patients planned to be recruited in total, about 180 had already been recruited, implying that in case of stopping the trial at the IA, 440 - 180 = 260 would not be recruited. Taking into account this saving in patient recruitment and comparing the 15 months to the time until final analysis as provided in Table~\ref{tab:MIRROSsize}, 38.8 months, the estimated IA timepoint to make a decision on futility of the trial appears sensible. Operationally, the trial was initiated recruiting patients in a limited number of sites. More sites were opened only after passing the IA. Enrolment into the trial was not suspended between the IA cut-off date (defined by the first 120 recruited patients) and the iDMC recommendation, and as a consequence the data of about 60 patients (40 on experimental and 20 on control arm) was reviewed as part of a regular safety assessment, but not used for the futility decision. An efficient read-out process was set up to minimize the time for a futility decision, thus supporting continuation of recruitment without interruption. This is different to e.g. Hunsberger et al.\cite{hunsberger_09} who assume that recruitment is halted between read-out and when a final decision on stopping or continuation of the trial has been made.

Potential safety risks associated with administration of the new treatment were mitigated by holding regular safety review meetings by the iDMC.

As Chen et al.\cite{chen_18} point out, a randomized Phase 3 trial commands substantial upfront investment, and they comment that as a consequence, there is often little incentive to stop the trial midway for futility. As a result, the futility bar is often set low (meaning that already a small or even no effect suffices to pass the futility IA), rendering the IA nothing more than a ``disaster check'', or ``data cleaning exercise''. A low bar Phase 3 futility analysis can thus hardly replace a bona fide Phase 2 proof-of-concept trial. For MIRROS, neither of these concerns apply: the futility bar is high as illustrated by the quite high false-negative probability of the chosen $x^*$ in Section~\ref{opchar_IA}.

A key hurdle in integrating Phase 2 into Phase 3 is that the sponsor needs to agree to:
\bi
\item Increased upfront investment and planning to enable the Phase 3 design to be specified already at the outset of Phase 2.
\item Pivotal, inferentially seamless studies, as MIRROS, imply that the Sponsor does not have access to Phase 2 data. This limits the ability to reflect new information in the Phase 3 portion of the trial, or to inform other development programs in similar indications and/or using similar treatment approaches.
\ei
These issues are discussed e.g. by Cuffe et al.\cite{cuffe_14}.

In Section~\ref{RCTplanning} the dependence of power on the shape of the actual survival functions, recruitment, and dropout was discussed. Furthermore, while assumptions are informed by historical data as much as possible, they are estimates and thus come with uncertainty. We thus agree with the recommendation given in Hunsberger et al.\cite{hunsberger_09} that when designing a trial like MIRROS, various sets of parameters should be evaluated. Furthermore, when approaching the final analysis, it might be indicated to verify that with the actual accrual and dropout, power at the pre-specified number of events is still approximately what was initially targeted. Note that the dependence of power on assumptions on the HR in the uncured patients and the cure proportions is not monotone in increasing effect sizes, i.e. the power is really determined through the assumption on the entire survival curves in both groups, see Wang et al.\cite{wang_12}, Section 4.2.

\subsection{Current status and planned final analysis}
\label{mirrosstatus}

The MIRROS trial passed the futility IA in mid 2017 after the iDMC issued a positive recommendation based on its review of unblinded futility IA data.

The final analysis will take place once the necessary number of OS events to achieve the targeted power are observed. As discussed in Section~\ref{trial_success}, the trial will meet its primary endpoint if the null hypothesis \eqref{h0} for OS is rejected using an unweighted logrank test. Treatment-specific survival functions will be nonparametrically estimated using Kaplan-Meier, and the treatment effect will be quantified using the overall hazard ratio. To further summarize OS, median OS and yearly survival probabilities, all including 95\% confidence intervals, will be estimated. The latter will give an indication about cure proportions.

\section{Potential alternative designs}
\label{alternatives}

A possible alternative for the MIRROS design would have been to use an adaptive design to integrate Phase 2 into Phase 3. In such an adaptive design, efficacy results after 120 patients would have been analyzed and based on these, pre-specified adaptations for the post-IA part could have been implemented. In terms of analysis this would entail that results from pre- and post-IA would have to be combined using methodology for adaptive clinical trials, e.g. by a suitable $p$-value combination test\cite{wassmer_16}. First, note that in MIRROS no adaptation beyond potentially setting the sample size to 0 after the futility IA (which is sometimes also considered an ``adaptation'') was anticipated. Second, the relatively high futility bar in MIRROS implied a high confidence in the experimental arm once the futility IA is passed, not warranting any modification of the trial. Third, the intermediate endpoint of response in MIRROS is at best indicative of efficacy, it cannot be considered a full surrogate at this point on which to base an early efficacy assessment in an adaptive design. A recent discussion of these aspects is provided by Freidlin et al.\cite{freidlin_18}.

This last point also offers part of the explanation why no efficacy interim was added to MIRROS. In addition, the arguments why the futility interim decision was not based on OS outlined in Section~\ref{mirros} also apply to a potential IA for efficacy.

Instead of powering the trial to detect an assumed alternative HR one could also have considered setting up a hypothesis test comparing the cure proportions only. Or, more precisely, perform a hypothesis test that assesses the null hypothesis
\bean
  H_0 \ : \ S_1(t_0) = S_2(t_0) \label{h0milestone}
\eean for a suitably large $t_0$, i.e. a milestone timepoint at which patients can be considered cured with sufficient confidence. However, such a test would generally have less power compared to an unweighted logrank test for \eqref{h0}, at least in the range of scenarios considered in this paper, i.e. those induced through the cure proportion model from Section~\ref{model} with low to moderate cure proportions. For illustration, we have computed the power to test the null hypothesis \eqref{h0milestone} using the test statistic $X_3$ in Klein et al.\cite{klein_07} for $t_0 = 12, 18, 24, 30, 36$ months for the MIRROS scenario (Row 1) from Table~\ref{tab:MIRROSsize}. As expected, with values of 0.75, 0.73, 0.62, 0.44, 0.18, respectively, the power at these milestones is much lower than the 0.85 assumed to derive the necessary number of 275 events to detect the assumed alternative in Table~\ref{tab:MIRROSsize}.
In addition to the power reduction when looking at a milestone only, powering the trial to detect a difference at a milestone would also not capture a potential treatment effect of delaying deaths for those patients not cured and would require a definition of a timepoint beyond which patients would be considered cured.

Finally, Wang et al.\cite{wang_12} derive sample size formulas for alternatives built through all possible combinations of hypotheses on the survival functions of the uncured and cure proportions. These could have been used for sample size computation, but integration of Phase 2 using CR would not have been possible with that approach.


\section{Conclusions}
\label{discussion}

In this paper, we describe a clinical trial design that has two key features:
\bi
\item With OS as primary endpoint, it accounts for a cure proportion in both arms. The necessary number of events to get a pre-specified power is computed via simulations.
\item Based on a mechanistic simulation model that connects response and OS, the design integrates Phase 2 into Phase 3, i.e. has a futility IA based on response. A simulation model is used to compute operating characteristics of the interim decision under various assumptions.
\ei
The design allows to answer the key clinical question, namely whether addition of Idasanutlin to Cytarabine is able to improve OS in R/R AML, thereby accelerating the development program, in line with Chen et al.\cite{chen_18}, who consider it ``...imperative to improve the efficiency of drug development via innovation.'' However, the proposed design appropriately mitigates the risk of going from Phase 1 directly to Phase 3 using a futility IA based on an intermediate endpoint.

While not entirely standard and requiring simulation in the planning phase, the proposed design is still sufficiently transparent and pragmatic to get internal and external support. As discussed in Section~\ref{mirrosstatus}, the MIRROS trial has already passed the IA, indicating that CR proportions are higher for Idasanutlin plus Cytarabine vs. Cytarabine plus placebo. Whether this CR increase will translate into a survival benefit will only be known when the study reaches the final analysis.

The paper also illustrates practical challenges that need to be taken into account when implementing a design that appears straightforward in theory, e.g. the inability to identify and exclude the mutants at randomization.

\section{Software}
\label{software}

The code and the results of all computations described in this paper are available as a github repository: \url{https://github.com/numbersman77/integratePhase2.git}

\section{Acknowledgments}
\label{ack}

We thank Marcel Wolbers, Carol Ward, Olivier Catalani, Susanne M\"uhlbauer, Rachel Rosenthal, and Marion Ott for reviewing earlier versions of this paper and the entire Idasanutlin team for the collaboration.

\appendix

\section*{Appendix}

\section{Monthly and yearly drop-out rate}

Often in trial planning, a drop-out rate per year, $\delta$, is specified in protocols to allow to model events such as patient withdrawal from the trial, or patients' loss to follow-up. In order to have all rates on the same time-scale (months), this yearly rate can easily be transformed into a monthly rate $\tau$ by solving $1-\delta = (1-\tau)^{12}$ for $\tau$, which gives $\tau = 1-(1-\delta)^{1/12}$.

\section{Random number from the cure proportion model in Section~\ref{model}}

To obtain a random number from a distribution with survival function $S_i$, one can use $F_{i}^{-1}(U)$, where $U$ is a uniform random variable and
\bea
  F_{i}^{-1}(q) &=& (F_{i}^*)^{-1}\Bl(1-\frac{q-p}{1-p}\Br)
\eea the quantile function of $S_i$, for a quantile $q \in [p, 1]$ and $F_{i}$ and $F_{i}^*$ being the cumulative distribution functions corresponding to $S_{i}$ and $S_{i}^*$.

\bibliographystyle{ieeetr}
\bibliography{stat}

\end{document}

%% file: commands.tex
\def\bea{\begin{eqnarray*}}
\def\eea{\end{eqnarray*}}
\def\bean{\begin{eqnarray}}
\def\eean{\end{eqnarray}}
\def\bi{\begin{itemize}}
\def\ei{\end{itemize}}

\def\intb{x^*}

\usepackage{amsmath}
\usepackage{amssymb}
\usepackage{amsfonts}
\usepackage{amsthm}
\usepackage{latexsym}
\usepackage{color}
\usepackage{epsfig}
\usepackage{hyperref}
\usepackage{longtable}
\usepackage{multirow}
\usepackage{lscape}
\usepackage[normalem]{ulem}   

\def\Bl{\Bigl}
\def\Br{\Bigr}